\documentclass[12pt]{article}
\usepackage[]{epsfig}
\usepackage[centertags]{amsmath}
\usepackage{amsfonts}
\usepackage{amssymb}
\usepackage[english]{babel}
\usepackage{amsmath, amsthm, slashed,url}
\usepackage{float}

\begin{document}

\title{\bf A note on vortices from Lorentz-violating models }
\author{
Lucas Sourrouille$^{a,b}$
\\
{\normalsize \it $^a$Departamento de F\'\i sica, Universidade Federal do Maranh\~ao,
}\\ {\normalsize\it 65085-580, S\~ao Lu\'is, Maranh\~ao, Brazil}
\\
{\normalsize\it $^b$CMCC, Universidade Federal do ABC, Santo Andr\'e, SP, Brazil}
\\
{\footnotesize lsourrouille@yahoo.es } } \maketitle

\abstract{We consider two self-dual abelian Higgs systems obtained from Lorentz breaking symmetry
models by dimensional reduction. For the first model, we show that the self-dual equations are identical to 
those of Nielsen-Olesen vortices. Also, we show that our vortices have electric charge. In the second case we show
that self-dual Chern-Simons-Higgs vortices without electric charge are possible. }

\vspace{0.5cm}

{\bf pacs}: 11.10.Kk, 11.10.Lm, 11.27.+d, 12.60.Ài, 11.30.Cp


\vspace{1cm}
\section{Introduction}
The possible violation of Lorentz invariance has recently received a lot of attention as a candidate for the Planck scale
physics \cite{K1}.The large range of existing phenomenological and experimental activities stems from the application
of effective field theory\cite{K2} and the construction of the Standard-Model Extension (SME)\cite{K3,K4} to catalogue and
predict observable effects.
The gauge sector of the SME include CPT-even and CPT-odd terms. The CPT-odd, usully called the Carroll-Field-Jackiw
term\cite{CFJ}, consist on four parameters, that engenders a parity-odd and birefringent
electrodynamics, coupled to Chern-Simons dynamics. The CPT-even photon sector is composed by fourth-rank tensor,
with Riemann tensor symmetries and a double null trace, which is coupled to a gauge term $F^{\kappa \lambda } F^{\mu \nu }$. 
These models have been studied in different contexts, such as radiative corrections \cite{4},
nontrivial spacetime topology \cite{5}, causality \cite{6}, supersymmetry
\cite{8}, the cosmic microwave background \cite{10}, vacuum Cherenkov radiation\cite{cr}, general relativity \cite{11} and
soliton solutions, like monopoles\cite{mon} and vortices\cite{12,121,122}.
\\[3mm]
On the other hand, it is well known  that the abelian Higgs model \cite{NyO} in $(2+1)$ dimensions,
support topological vortex solutions of finite energy.  With a
specific choice of coupling constants, minimum energy vortex configurations
satisfy first order differential equations \cite{Bogo}. This model admits and $N=2$ supersymmetric
extension  providing a connection between BPS equations and supersymmetry \cite{susy1}.
Also the $(2+1)$-dimensional matter field interacting with gauge fields
whose dynamics is governed by a Chern–Simons term support soliton solutions \cite{Jk}. When self-interactions are suitably 
chosen, 
vortex configurations satisfy the Bogomol’nyi-type equations with a specific sixth order potential \cite{JW}. Another important 
feature of the Chern–Simons gauge field is that inherits its dynamics from the matter fields to which it is coupled, so it may be 
either relativistic \cite{JW} or non-relativistic \cite{JP}. In addition the soliton solutions
are of topological and non-topological nature \cite{JLW}.
\\[3mm]
The purpose of this letter is the study of the self-dual vortices in two Lorentz-violating models. In the first section we will
analyze the self-dual equations of a two dimensional model obtained from the $CPT$-even Lorentz-violating term, coupled to
matter field, by dimensional reduction. We will show that this model support self-dual vortex solutions with electric charge.
In particular, we will show that the self-duality equations are identical to those of the Nielsen-Olesen vortices.
The second section is dedicated to explore the dimensional reduction of the $CPT$-odd Lorentz-violating term couple to a Higgs
field. We will show, here, that it is possible to obtain exactly the same Bogomolnyi equations as the
relativistic Chern-Simons-Higgs model. The novelty is that our vortices have no electric charge unlike the
Chern-Simons-Higgs vortex solutions.

\section{Nielsen-Olesen vortices with electric charge from $CPT$-even Lorentz-violating model}

Let us start by considering the $(3+1)$-dimensional $CPT$-even Lorentz-violating model governed by the action \cite{K3},
\begin{equation}
S =\int d^{4}x\Big( -\frac{1}{4} k_{\kappa \lambda \mu \nu } F^{\kappa \lambda } F^{\mu \nu } -A^{\nu }J_{\nu
}\Big)\;, \label{Ac1}
\end{equation}
where the metric tensor is $g^{\mu \nu }=(1,-1,-1,-1)$.
The coupling $k_{\kappa \lambda \mu \nu }$ is real and dimensionless. It has the symmetries of the Riemann tensor and zero
double trace, so it
contains $19$ independent real components. The field strength $F^{\mu \nu }$ is defined by
\begin{equation}
F^{\mu \nu}=\partial^{\mu}A^{\nu}-
\partial^{\nu}A^{\mu}
\end{equation}
and the term $A^{\nu
}J_{\nu }$ represents the coupling between the gauge field and an external
current.

The equation of motion arising from the Lagrangian (\ref{Ac1}) is
\begin{equation}
k_{ \mu \alpha \beta \gamma}\partial^{\alpha }F^{\beta \gamma} +J^{\mu }=0
\label{EqM}
\end{equation}
In the following we will consider $(3+1)$ dimensional $CPT$-even Lorentz-violating model coupled to Higgs ($\phi$)
matter
\begin{eqnarray}
S=\int d^{4}x\Big( -\frac{1}{4} k_{\kappa \lambda \mu \nu } F^{\kappa \lambda } F^{\mu \nu } +|D_{\mu }\phi |^{2}-V(|\phi|)
\Big)
\label{Ac2}
\end{eqnarray}
The covariant derivative is defined as $D_{\mu}= \partial_{\mu} + ieA_{\mu}$ and $V(|\phi|)$ is a self-interacting potential.
Since, we will analyze the dimensional reduction of the model (\ref{Ac2}), it is convenient to expand the first term of the
action,
\begin{eqnarray}
S_{CPT} &=&-\frac{1}{4}\int d^{4}x  k_{\kappa \lambda \mu \nu } F^{\kappa \lambda } F^{\mu \nu }\nonumber \\
& =& -2 \int d^{4}x \Big( k_{0 i 0
i } (F^{0i})^2 + k_{i j i j} (F^{ij})^2 + k_{0 j 0 i} F^{0j}F^{0i} +  \nonumber \\&& k_{1 \mu 1 \nu} F^{1\mu} F^{1\nu}
+ k_{2 \mu 2 \nu} F^{2\mu} F^{2\nu} + k_{3 \mu 3 \nu} F^{3\mu} F^{3\nu} + k_{0 i j \kappa} F^{0 i}
F^{j \kappa} \Big)\;,
\end{eqnarray}
where
\begin{eqnarray}
k_{0 i 0 i }(F^{0i})^2 &=& k_{0 1 0 1 }(F^{01})^2 + k_{0 2 0 2 }(F^{02})^2 + k_{0 3 0 3 }(F^{03})^2
\nonumber \\[3mm]
k_{i j i j} (F^{ij})^2 &=& k_{1212 }(F^{12})^2 + k_{1313}(F^{13})^2 + k_{2323 }(F^{23})^2
\nonumber \\[3mm]
k_{0 j 0 i} F^{0j}F^{0i} &=& k_{0102}F^{01}F^{02} + k_{0103}F^{01}F^{03}+ k_{0203 }F^{02}F^{03}
\nonumber \\[3mm]
k_{1 \mu 1 \nu} F^{1\mu}F^{1\nu} &=& k_{1012}F^{10}F^{12} + k_{1013}F^{10}F^{13}+ k_{1213 }F^{12}F^{13}
\nonumber \\[3mm]
k_{2 \mu 2 \nu} F^{2\mu}F^{2\nu} &=& k_{2023}F^{20}F^{23} + k_{2021}F^{20}F^{21}+ k_{2123 }F^{21}F^{23}
\nonumber \\[3mm]
k_{3 \mu 3 \nu} F^{3\mu}F^{3\nu} &=& k_{3031}F^{30}F^{31} + k_{3032}F^{30}F^{32}+ k_{3132 }F^{31}F^{32}
\nonumber \\[3mm]
k_{0 i j \kappa} F^{0i}F^{j\kappa} &=& 2k_{0123}F^{01}F^{23} +2 k_{0312}F^{03}F^{12}+ 2k_{0231}F^{02}F^{31}
\label{EqM2}
\end{eqnarray}
We will study the soliton solution in a $(2+1)$-dimensional model, obtained from
(\ref{Ac2}) via dimensional reduction. In order
to carry out this procedure it is natural to consider a
dimensional reduction of the action by suppressing dependence on the third spacial coordinate and renaming $A_3$ as $N$. Then,
the action (\ref{Ac2}) becomes
\begin{eqnarray}
S_{(2+1)}=\int d^{3}x\Big( S_{rCPT} +|D_0 \phi |^{2} - |D_{i}\phi |^{2} -e^2 N^2|\phi|^2 -V(|\phi|)
\Big)\;,
\label{Ac3}
\end{eqnarray}
where $S_{rCTP}$ is the reduced $CPT$-even Lorentz-violating action,
\begin{eqnarray}
S_{rCPT}&=& -2\int d^{3}x\Big( k_{0i0i} (F^{0i})^2 + k_{\mu 3 \mu 3} (\partial^\mu N)^2 + k_{1212} (F^{12})^2 +  k_{\mu \nu \mu
3}F^{\mu \nu} \partial^\mu N
\nonumber \\[0.2cm]&&
+ k_{0312} F^{12} (\partial^0 N) + k_{0ij3} F^{0i} (\partial^j N) + k_{i0ij} F^{i0} F^{ij}
\nonumber \\[0.2cm]&&
+ k_{0102} F^{01} F^{02} + k_{303i} (\partial^0 N)(\partial^i N) + k_{3132} (\partial^1 N)(\partial^2 N) \Big)\;,
\label{Ac4}
\end{eqnarray}
with $(i,j =1,2)$ and $(\mu,\nu =0,1,2)$.

Here, we are interested in time-independent soliton solutions that ensure the finiteness of the action (\ref{Ac3}).
These are the stationary points of the energy which for the static field configuration reads
\begin{eqnarray}
E_{(2+1)}&=& \int d^{2}x \Big[2\Big( k_{0i0i} (\partial^{i}A^0)^2 + k_{i 3 i 3} (\partial^i N)^2 + k_{1212} (F^{12})^2 +  k_{i
j i 3}
F^{ij} \partial^i N
\nonumber \\[0.2cm]&&
+k_{i0i3} \partial^i A^0 \partial^i N + k_{0ij3} (-\partial^i A^0) (\partial^j N) + k_{i0ij}  (\partial^i A^0) F^{ij}
\nonumber \\[0.2cm]&&
+ k_{0102} (\partial^1 A^0) (\partial^2 A^0) + k_{3132} (\partial^1 N)(\partial^2 N) \Big)+ e^2|\phi|^2(N^2 - A^2_0)
\nonumber \\[0.2cm]&&
+ |( D_1 \pm iD_2)\phi|^2 \mp eF_{12}|\phi|^2 \pm \frac{1}{2} \epsilon^{ij} \partial_i J_j + \frac{\lambda}{4} (|\phi|^2
-v^2)^2\Big]\;,
\label{Ac5}
\end{eqnarray}
where we have used the fundamental identity
\begin{eqnarray}
|D_i \phi|^2 = |( D_1 \pm iD_2)\phi|^2 \mp eF_{12}|\phi|^2 \pm \epsilon^{ij} \partial_i J_j
\label{iden}
\end{eqnarray}
The potential term has been chosen so that it coincides with the symmetry breaking potential of the Higgs model
\begin{eqnarray}
V(|\phi|)=\frac{\lambda}{4} (|\phi|^2-v^2)^2
\label{pot}
\end{eqnarray}
The term $\int d^{2}x\frac{1}{2} \epsilon^{ij} \partial_i J_j$ is a surface term and may be dropped with the hypothesis that
\begin{eqnarray}
\phi|_{|x|=\infty} = \upsilon e^{i\alpha (\theta)}
\nonumber \\[3mm]
A_i|_{|x|=\infty} = \frac{i}{e} \partial_i (\log \phi)
\label{Co1}
\end{eqnarray}
This condition also implies that the covariant derivative must vanish
asymptotically, which fixes the behavior of the gauge field $A_i$. Then we have
\begin{eqnarray}
\Phi = \int \,\,d^2 x F_{12} = \oint_{|x|=\infty} \,\, A_i dx^i  = 2\pi n
\end{eqnarray}
where $n$ is a topological invariant which takes only integer values.

The formula (\ref{Ac5}) may be rewritten as a sum of square terms which are
bounded below by a multiple of the magnitude of the magnetic flux. To achieve this we can regroup the terms of
(\ref{Ac5}) as follows,
\begin{eqnarray}
k_{0i0i} (\partial^{i}A^0)^2 + k_{i 3 i 3} (\partial^i N)^2 +k_{i0i3} \partial^i
A^0 \partial^i N =k_{0i0i} (\partial^i A^0 \pm \partial^i N)^2\;,
\end{eqnarray}
where, we have chosen $k_{0i0i}=k_{i 3 i 3}=\pm \frac{1}{2}k_{i0i3}$. In addition, if we choose $\pm k_{i0ij}=k_{iji3}$, we have
\begin{eqnarray}
k_{iji3}F^{ij} \partial^i N + k_{i0ij}  \partial^i A^0 F^{ij} = k_{i0ij} F^{ij} (\partial^i A^0 \pm \partial^i N)
\end{eqnarray}
Since
\begin{eqnarray}
k_{0ij3} (-\partial^i A^0) (\partial^j N)= -k_{0123} (\partial^1 A^0) (\partial^2 N) + k_{0213} (\partial^2 A^0) (\partial^1 N)\;,
\end{eqnarray}
we can choose, $-k_{0123}=\pm k_{3132}$ and $k_{0102}=\pm k_{0213}$, so that
\begin{eqnarray}
k_{3132} (\partial^1 N) (\partial^2 N)-k_{0123} (\partial^1 A^0) (\partial^2 N) = k_{0123} \partial^2 N(-\partial^1 A^0 \mp
\partial^1 N)
\nonumber \\[3mm]
k_{0213} (\partial^2 A^0) (\partial^1 N) + k_{0102} (\partial^2 A^0) (\partial^1 A^0) = k_{0102} \partial^2 A^0(\partial^1 A^0
\pm \partial^1 N)
\end{eqnarray}
Finally, we can write the following identity:
\begin{eqnarray}
&&2k_{1212}(F^{12})^2 \mp eF^{12}|\phi|^2 + \frac{\lambda}{4} (|\phi|^2-v^2)^2 =
\nonumber \\[0.2cm]&&
\frac{\kappa}{2}[F^{12} \mp
\frac{e}{\kappa}(|\phi|^2 -v^2)]^2 +
(2k_{1212} -\frac{\kappa}{2})(F^{12})^2 +
\nonumber \\[0.2cm]&&
(\frac{\lambda}{4}
-\frac{e^2}{2\kappa})(|\phi|^2-v^2)^2 \mp ev^2 F^{12}\label{r1}
\end{eqnarray}
If the coupling constant $\lambda$  and $\kappa$ are chosen such that
\begin{eqnarray}
\lambda= \frac{e^2}{2k_{1212}}
\nonumber \\[3mm]
\kappa= 4k_{1212}\;,
\end{eqnarray}
the expression (\ref{r1}) takes the simple form
\begin{eqnarray}
&&2k_{1212}(F^{12})^2 \mp eF^{12}|\phi|^2 + \frac{\lambda}{4} (|\phi|^2-v^2)^2 =
\nonumber \\[0.2cm]&&
2k_{1212}[F^{12} \mp
\frac{e}{4k_{1212}}(|\phi|^2 -v^2)]^2 \mp ev^2 F^{12}
\end{eqnarray}
Using these identities, the energy functional (\ref{Ac5}) becomes
\begin{eqnarray}
E_{(2+1)}&=& \int d^{2}x \Big[2\Big(k_{0i0i} (\partial^i A^0 \pm \partial^i N)^2 +k_{i0ij} F^{ij} (\partial^i A^0 \pm \partial^i
N)
\nonumber \\[0.2cm]&&
+k_{0123} \partial^2 N(-\partial^1 A^0 \mp\partial^1 N) +k_{0102} \partial^2 A^0(\partial^1 A^0\pm \partial^1 N)
\nonumber \\[0.2cm]&&
+k_{1212}[F^{12}\mp \frac{e}{4k_{1212}}(|\phi|^2 -v^2)]^2 \Big)+  e^2|\phi|^2(N^2 - A^2_0)
\nonumber \\[0.2cm]&&
+|( D_1 \pm iD_2)\phi|^2\mp ev^2 F^{12} \Big]
\label{Ac6}
\end{eqnarray}
If we choose $A_0=\mp N$, then the fields $A_0$ and $N$ are decoupled to the fields $A_i$ and $\phi$ and we obtain an expression
for the energy similar to the well known energy for the Maxwell-Higgs model
\begin{eqnarray}
E_{(2+1)}= \int d^{2}x \Big(
2k_{1212}[F^{12}\mp \frac{e}{4k_{1212}}(|\phi|^2 -v^2)]^2
+|( D_1 \pm iD_2)\phi|^2\mp ev^2 F^{12} \Big)\;,
\label{Ac6}
\end{eqnarray}
Thus, the energy (\ref{Ac6}) is reduced
to a sum of square terms which are bounded below by a multiple of the magnitude of the magnetic flux(for positive
flux we choose the lower signs, and for negative flux we choose
the upper signs):
\begin{eqnarray}
E_{(2+1)} \geq ev^2 |\Phi|
\end{eqnarray}
The bound is saturated by fields satisfying the first-order Bogomolnyi \cite{Bogo} self-duality equations:
\begin{eqnarray}
& &( D_1 \pm iD_2)\phi =0
\nonumber \\
& &
F^{12} \mp \frac{e}{4k_{1212}} (|\phi|^2 -v^2)=0\;,
\label{bogo1}
\end{eqnarray}
The equations (\ref{bogo1}) are identical in form to the Bogomolnyi equations of the abelian Maxwell-Higgs model. The difference 
lies in the fact that our vortices not only carry magnetic flux, as in the Higgs model, but also electric charge. This is a 
consequence that in our theory
the
dynamics of gauge field is dictated by a $CPT$-even Lorentz-violating term instead of a Maxwell term as in Higgs theory.

\section{Self-Dual Chern-Simons vortices without electric charge from $CPT$-odd Lorentz-violating model}
In this section, we analyze the $CPT$-odd Lorentz-violating model \cite{CFJ}, coupled to Higgs field
\begin{eqnarray}
S &=&\int d^{4}x\Big[\frac{1}{4}%
p_{\alpha }\epsilon ^{\alpha \beta \mu \nu }A_{\beta }F_{\mu \nu }
+|D_{\mu }\phi |^{2}+V(|\phi |)\Big],
\label{Ac7}
\end{eqnarray}
where the coupling coefficient  $p_\alpha$ is real and has dimensions of mass. The antisymmetric tensor $\epsilon ^{\alpha \beta 
\mu \nu }$ is the totally antisymmetric tensor such that $\epsilon^{0123}=1$.

By varying with respect to $A_3$ and $A_0$, we obtain the field equations
\begin{equation}
p_{\alpha}\frac{1}{2}\epsilon ^{ \alpha 3 \mu \nu
}F_{\mu \nu } + J_{3 }=0\;,
\label{EqM3}
\end{equation}

\begin{equation}
p_{\alpha }\frac{1}{2}\epsilon ^{\alpha 0 \mu \nu
}F_{\mu \nu } + J_{0 }=0\;,
\label{EqM4}
\end{equation}
where $J_0 = -ie\Big[\phi^*(D_0 \phi) - \phi(D_0 \phi)^* \Big]$ and $J_3 = ie\Big[\phi^*(D_3 \phi) - \phi(D_3 \phi)^* \Big]$.

Here, we are interested in the analysis of the structure of first-order Bogomol’nyi self-duality equations for $p_i=0$ $(i
=1,2,3)$ and $p_0\not=0$. This condition implies a reduction of the action (\ref{Ac7}) and the equations (\ref{EqM3}), 
(\ref{EqM4})

\begin{eqnarray}
S &=&\int d^{4}x\Big[\frac{1}{4}%
p_{0 }\epsilon ^{0 \beta \mu \nu }A_{\beta }F_{\mu \nu }
+|D_{\mu }\phi |^{2}+V(|\phi |)\Big],
\label{Ac8}
\end{eqnarray}
\begin{eqnarray}
p_{0}F_{1 2 } + J_{3 }=0\;,
\label{EqM5}
\end{eqnarray}
\begin{eqnarray}
J_{0 }=0
\label{EqM6}
\end{eqnarray}
We can proceed similarly to previous section and reduce one dimension by assuming that the fields do not
depend on one of the spatial coordinates, say  $x_3$, and renaming $A_3$ as $N$,
\begin{eqnarray}
S_r =\int d^{3}x\Big(\frac{1}{4}p_{0}(2 A_1 \partial_2 N -2A_2 \partial_1 N +2N F_{12})\nonumber \\
+|D_0 \phi |^{2} - |D_{i}\phi |^{2} -e^2 N^2|\phi|^2 +V(|\phi|)\Big)\;,
\label{}
\end{eqnarray}
Integrating by parts the two first terms of this action, we have, 
\begin{eqnarray}
S_r =\int d^{3}x\Big(
 Np_{0 }F_{1 2} +|D_0 \phi |^{2} - |D_{i}\phi |^{2} -e^2 N^2|\phi|^2 +V(|\phi|)\Big)\;,
\label{Ac9}
\end{eqnarray}
The equation (\ref{EqM5}) becomes
\begin{equation}
 N = \frac{p_{0}F_{1 2 }}{2e^2 |\phi|^2}\,,
\label{EqM7}
\end{equation}
Substitution of this equation into the expression of the action (\ref{Ac9}) yields  
\begin{eqnarray}
S_r =\int d^{3}x\Big(
 \frac{p_0^2 F^2_{12}}{4e^2|\phi|^2} +|D_0 \phi |^{2} - |D_{i}\phi |^{2} +V(|\phi|)\Big)\;,
\label{Ac10}
\end{eqnarray}
For the static field configuration the equation (\ref{EqM6}) reads
\begin{eqnarray}
2e^2 A_0 |\phi|^2=0
\label{EqM8}
\end{eqnarray}
Thus, the action (\ref{Ac10}) may be rewritten as
\begin{eqnarray}
S_r =\int d^{2}x\Big(
 \frac{p_0^2 F^2_{12}}{4e^2|\phi|^2} - |( D_1 \pm iD_2)\phi|^2 \pm eF_{12}|\phi|^2 +V(|\phi|)\Big)\;,
\label{Ac11}
\end{eqnarray}
where the identity (\ref{iden}) was used .

To uncover the Bogomol’nyi-style self-duality, the action is expressed in the form 
\begin{eqnarray}
S_r =\int d^{2}x&\Big(&\Big[\frac{ p_0}{2e |\phi|} F_{12} \mp \frac{e^2}{ p_0} |\phi|(v^2 -|\phi|^2)\Big]^2 + |i( D_1 \pm
iD_2)\phi|^2
\nonumber \\[0.2cm]
&&\pm e v^2 F_{12} +V(|\phi|) - \frac{e^4}{p_0^2} |\phi|^2(v^2 -|\phi|^2)^2\Big)\;,
\label{}
\end{eqnarray}
Thus, if the potential is chosen to take the self-dual form
\begin{eqnarray}
V(|\phi|) = \frac{e^4}{p_0^2} |\phi|^2(v^2 -|\phi|^2)^2\;,
\end{eqnarray}
the action is bounded below [choosing signs depending on the sign of the flux]
\begin{eqnarray}
S_{r} \geq ev^2 |\Phi|
\end{eqnarray}
Therefore, when the action reaches its minimum value, the fields will extremise the action and satisfy the 
first-order field equations 
\begin{eqnarray}
& &( D_1 \pm iD_2)\phi =0
\nonumber \\
& &
F_{12} =\pm \frac{2e^3|\phi|^2}{p_0^2} (v^2 -|\phi|^2)\;,
\label{bogo2}
\end{eqnarray}
These equations may be compared with the self-duality equations of the Chern-Simons Higgs theory. If we
choose $p_0$ as the Chern-Simons coupling parameter $\kappa$, the equations (\ref{bogo2}) coincide with the well know
Bogomolnyi equations of the relativistic Chern-Simons-Higgs model. Nevertheless, the vortex solutions of the equation
(\ref{bogo2}) are not the identical to those of the relativistic Chern-Simons-Higgs theory. This is because the electric charge
density for our solitons is zero, as shown in equation (\ref{EqM6}).
\\
Finally, it is interesting to note, that in our model the 
Chern-Simons field does not lead to the equation of the type $E_i\propto \epsilon_{ij}J^j$, which is an equation of the 
Chern-Simons-Higgs theory. This is due to that we have considered $p_i=0$. 
\\
Another difference with the self-dual Chern-Simons-Higgs vortices lies in the fact that in our model the charge density is not 
locally proportional to the magnetic field, as shown the equations  (\ref{EqM5}) and (\ref{EqM6}). Thus, in the case of a static 
field configuration, we should infer the behavior of the field $A_0$ from the equation (\ref{EqM8}). The equation (\ref{EqM8}) 
shows that $A_0$ may be different from zero if and only if $\phi(x)=0$. Since, our vortex solutions are determined by equations 
(\ref{bogo2}), we know, from the Chern-Simons-Higgs solutions, that $\phi(x)$ has an only one root at $x=0$. Therefore, $A_0$ may 
be different from zero, only, at $x=0$. Nevertheless, in order to ensure the regularity of the electric field at $x=0$, we should 
impose that $A_0(0)=0$. Thus, our vortex solution has no electric field.

\vspace{0.6cm}
In summary we have discussed the Bogomolnyi framework for two models with Lorentz breaking symmetry. In the first case we have
analyzed  the dimensional reduction of a $CPT$-even Lorentz-violating term coupled to Higgs field,
showing that the model supports vortex solutions which are identical to those of Maxwell-Higgs model.
In the second situation we deal with a reduction of a Carroll-Field-Jackiw term coupled to matter. We have
shown, here, that the Bogomolnyi equations of
Chern-Simons Higgs theory are obtained as a particular case. It is important to note that in our first model, the vortex 
solutions has electric charge. This is a difference with the Nielsen-Olesen solutions. Our second model does not have electric 
charge, being this a difference with relativistic Chern-Simons-Higgs solitons. 
Finally, we want to note that vortices with electric charge in the 
CPT-even Lorentz-violating model coupled to matter, were found in Ref.\cite{121}. However the authors use a very 
strange 
potential term to generate these solutions, and the self-duality equations differ absolutely from our result.

\vspace{2cm}
{\bf Acknowledgements}\\
I would like to thank Rodolfo Casana for useful comments.
Also, I would like to thank Dmitri Vassilevich for hospitality during the realization
of this work. I am grateful to the referee for his careful reading of the manuscript. This work is supported by CAPES (PNPD/2011).

\end{document}